# CNN-based InSAR Denoising and Coherence Metric

Subhayan Mukherjee[1], Aaron Zimmer[2], Navaneeth Kamballur Kottayil[1], Xinyao Sun[1], Parwant Ghuman[2], Irene Cheng[1]

[1]Department of Computing Science, University of Alberta, Edmonton, Canada

[2]3v Geomatics, Vancouver, Canada

{mukherje,kamballu,xinyao1,locheng}@ualberta.ca      {azimmer,pghuman}@3vgeomatics.com

*Abstract*—Interferometric Synthetic Aperture Radar (InSAR) imagery for estimating ground movement, based on microwaves reflected off ground targets is gaining increasing importance in remote sensing. However, noise corrupts microwave reflections received at satellite and contaminates the signal's wrapped phase. We introduce Convolutional Neural Networks (CNNs) to this problem domain and show the effectiveness of autoencoder CNN architectures to learn InSAR image denoising filters in the absence of clean ground truth images, and for artefact reduction in estimated coherence through intelligent preprocessing of training data. We compare our results with four established methods to illustrate superiority of proposed method.

## I. INTRODUCTION

Remote sensing using activate microwave, especially in the form of Synthetic Aperture Radar Interferometry (InSAR), has been extensively used in decades. Highly accurate Digital Elevation Maps (DEMs) can be created by unwrapping the interferometric phase. However, the phase itself is often contaminated with noise arising from various sources, e.g. atmospheric factors. Hence, denoising the phase before unwrapping is essential to enhance the performance of the entire InSAR processing pipeline. The boxcar filtering approach is still widely used today. This essentially involves computing moving average using a rectangular window. However, ground images acquired using InSAR constantly change (also called "non-stationary") due to variations in topography and land displacements along the line of sight. This makes sample average methods like boxcar not the best solution for denoising InSAR images [1]. Also, strong smoothing effect of boxcar filtering results in spatial resolution loss, and noticeable errors in phase and coherence estimation near signal discontinuities. Over the past few decades, various filtering methods have been developed to address the problem of non-stationary InSAR phase estimation. They are broadly classified as spatial methods, e.g., Lee [2], and frequency based methods, e.g., Goldstein [3]. Both filters, as well as [4] are adaptive to the local fringe direction. The Lee filter averages similar phase values in the locally adaptive estimation window, while the Goldstein filter discards all but the most dominant component of the local power spectrum. Both the original and the modified Goldstein filter of Baran et al. [5] preserves the signal in high coherence (low noise) areas, and thus, are adaptive to local noise as well. Enhancements of the Lee filter [6-9] improve the adaptation to local fringe structure, whereas modifications to the Goldstein and Baran filters improve coherence estimation to avoid under-filtering the incoherent regions [10, 11]. Frequency based methods have been extended to the wavelet domain [12], including un-decimated wavelet transform [13] and wavelet packets [14], as filtering in wavelet domain preserves the spatial resolution. Local modeling based on polynomial approximation [15] and sparse coding [16] have been explored. Markov Random Field (MRF) based methods [17, 18] have also been attempted, though prior distribution modeling (required for MRF) is a challenging problem in itself. Recent attempts at Non-local filtering of various types of SAR images, which perform pixel estimation by matching patches over the whole image instead of just the pixel neighborhood, produced state-of-the-art results on amplitude, interferometric phase, polarimetric and multitemporal stack images [20, 21].

While Neural Network based despeckling of SAR images [22-25] and geo-localization accuracy improvement of optical satellite images [26] have been attempted, the use of Convolutional Neural Network (CNN) based learning approaches to InSAR images has not been explored. In this paper, we propose interferogram denoising and coherence prediction using two separate CNN architectures. We show:
1. In the absence of any clean ground truth, autoencoder CNN architecture is useful for unsupervised learning of denoising filters using only real-world noisy InSAR images.
2. Intelligent preprocessing of raw coherence can train the coherence estimation CNN to reduce various artefacts.

## II. PROPOSED METHOD

### A. Dimensionality Reduction for InSAR Denoising via CNN

While most supervised neural network based learning architectures require pairs of noisy and clean training images to learn how to transform an unseen noisy image to clean image, autoencoders do not have this limitation. Autoencoders can perform unsupervised learning as they can learn from the noisy data itself to reconstruct the input noisy data. Since it performs dimensionality reduction within the network, it learns to reconstruct from a latent representation of its input, which is in a lower dimensional space (the 'encoded' layer). Consequently, it learns only features that are essential to reconstruct its input, getting rid of the noise in the process. However, if there is excessive dimensionality reduction, the image detail also gets lost along with the noise. Thus, there is a trade-off.

The structure of our autoencoder is shown in Fig. 1. It should be noted that we do not constrain the size of the images used to train the network or run inference using the trained network (we can use whole image or patches). Also, the number of channels in the input is two, representing the real and imaginary part of the complex interferogram image. Thus, we exploit information from both channels. Nevertheless, we first process both the channels separately to saturate the outlier amplitudes, as some pixels in the input interferogram might have extremely high amplitudes. This may confuse the CNN during training / inference. Let the pixels in the interferogram be represented by $Z = [z_1, z_2, ... z_N]$ with amplitudes $A = [a_1, a_2, ... a_N]$. We can compute the amplitude of each pixel and threshold them to be $A' = $ saturate($A, M$), where $M$ is the mask denoting outlier amplitude values. The outliers are computed as in [27]. After saturation and normalization, real and imaginary channel values lie between -1 and +1. We add 1 to the two channels to use the Rectified Linear Unit (ReLU) activation for introducing nonlinearity in the CNN to learn complex features.

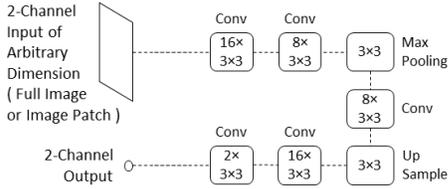

**Fig. 1**: Autoencoder CNN for denoising InSAR images.

In Fig.1, each CNN layer is represented by a box. Output feature map count is indicated by the integer at the top (2, 8, 16) whereas filter dimension is indicated at the bottom (3×3). Each 2D convolutional layer learns a number of filters. The maxpooling layer subsamples its input feature maps, while the upsampling layer brings them back to their original size. The output of the last convolutional layer has two feature maps representing the real and imaginary channel of the denoised interferogram. Mean squared error between the output channels and their corresponding input channels is reduced using the popular Adam optimizer to train the network, by updating its filter weights and biases, using gradient backpropagation. The network is trained using 60×60 patches extracted from noisy, real-world InSAR images. Maxpool downsampling factor is 3. Xavier style weight initialization is used for both networks.

*B. CNN-based Coherence Estimation to reduce Artefacts*

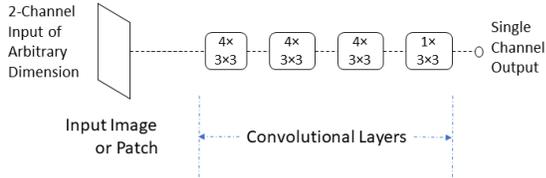

**Fig. 2**: CNN for InSAR coherence estimation.

We propose the coherence estimation CNN shown in Fig. 2. This architecture also has two channel noisy interferogram input, and the preprocessing for the channels is same as the one described earlier. However, we do not have any downsampling or upsampling, and the output is a single channel, as we require prediction of just the amplitude of the coherence. While the design is similar to the earlier network, the activation for final convolution layer is sigmoid instead of ReLU. This is because, pixel coherence is a value between 0 and 1 (sigmoid output). We use kernel regularization in the $2^{nd}$ last convolution layer to limit standard deviation between learned filter weights for that layer. This helps make the coherence estimates sharper and less grainy. For training this network, we first use 11×11 patches to compute the raw coherence between the training noisy images and their filtered version output by the fully trained filtering network described earlier. Raw coherence is defined in Eq. 1:

$$\hat{\gamma} = \frac{\sum_{n,m} u_1(n,m) \cdot u_2^*(n,m) \cdot e^{-j\phi(n,m)}}{\sqrt{\sum_{n,m} |u_1(n,m)|^2 \sum_{n,m} |u_2(n,m)|^2}} \quad (1)$$

where pixel $n$ of interferogram $u_1$ and pixel $m$ of interferogram $u_2$ have the angle of separation $\Phi$ and the asterisk on top of $u_2$ denotes complex conjugate. A relatively larger window size is used to reduce bias in raw coherence computation, but this also makes the coherence speckled. So, instead of training directly on raw coherence, we preprocess the raw coherence to make it sharper. We first segment the raw coherence using Chan-Vese segmentation [28], which is useful for segmenting objects with ill-defined boundaries like our case. The segmentation isolates incoherent areas from coherent ones. We set coherence of all coherent areas to 1 (full coherence). For each incoherent area,

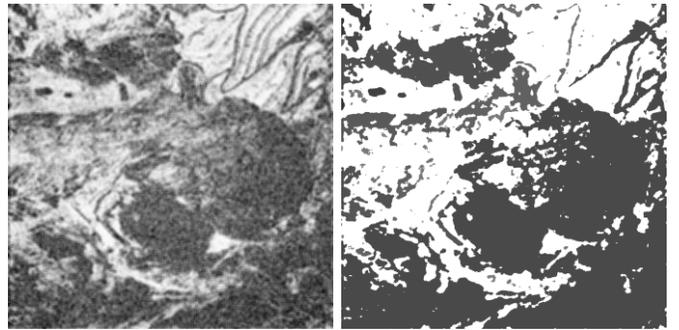

**Fig. 3**: Preprocess raw 11×11 coherence to remove artefacts.

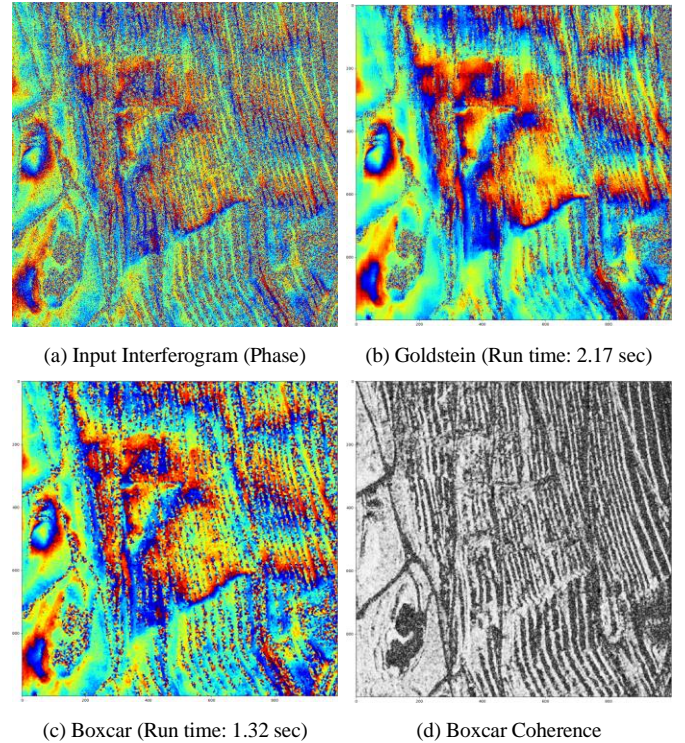

(a) Input Interferogram (Phase)    (b) Goldstein (Run time: 2.17 sec)

(c) Boxcar (Run time: 1.32 sec)    (d) Boxcar Coherence

Funding for this research was provided by MITACS and CARIC.

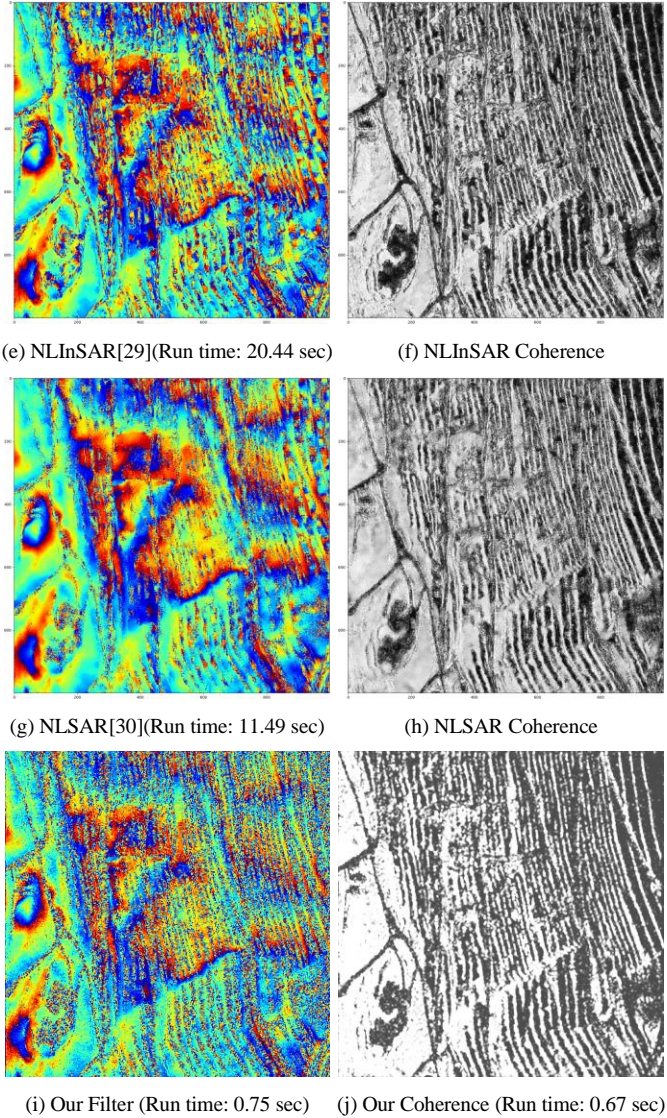

(e) NLInSAR[29](Run time: 20.44 sec)  (f) NLInSAR Coherence

(g) NLSAR[30](Run time: 11.49 sec)  (h) NLSAR Coherence

(i) Our Filter (Run time: 0.75 sec)  (j) Our Coherence (Run time: 0.67 sec)

**Fig. 4**: Filtering and coherence outputs of proposed method vs. established methods, colored between $-\pi$ (blue) to $+\pi$ (red) and 0 (black: low) to 1 (white: high) respectively.

we set its coherence to { Mean minus Standard Deviation } of coherence of all pixels comprising that area. We then extract corresponding 64×64 patches from training noisy image and its preprocessed coherence to train the coherence prediction CNN. The result of pre-processing is shown in Fig. 3 for a sample training image, where dark and light pixels show values closer to 0 (low coherence) and 1 (high coherence) respectively.

### III. RESULTS AND DISCUSSION

We implemented both the networks described earlier using Keras with Tensorflow back-end. We trained the filtering CNN and coherence estimation CNN by extracting 500 60×60 sized and 64×64 sized patches patches respectively from each of 135 1000×1000 training interferograms. The CNNs converged after 50 and 100 epochs respectively. The trained CNNs were tested on a set of 1000×1000 interferograms obtained from a different geographic location, but using the interferogram itself as input, instead of just patches. Fig. 4 compares our method's filtering, coherence and run time performance with 4 existing methods.

Fig. 4 shows that our method produces better demarcation between coherent and incoherent regions and far less spatial variance in coherence estimates in incoherent areas than all other methods, because we use a large window size (11×11) for raw coherence computation, preprocess it before training, and regularize kernel weights of the $2^{nd}$ last convolution layer. Also, our denoising does not have a tendency to flatten the phase in fully noisy areas, which is an artefact of the NLInSAR and NLSAR methods, though NLInSAR preserves details better. Goldstein and NLSAR tend to warp fringes and Goldstein under-filters near incoherent areas. Furthermore, the total execution time for our method (0.75 for denoising + 0.67 for coherence estimation = 1.42 seconds) is lower than most other methods. All methods were implemented and executed in OpenCL 1.2 on a NVIDIA 1070 GPU with 8 GB GPU RAM.

A potential improvement of our proposed method is to lower bias in raw coherence computation introduced by multiplying the input interferogram with complex conjugate of its filtered version. Our method intends to prevent the signals from biasing down the coherence estimates. However, where denoising fails to remove noise, this cancels out noise in areas where noise is present in both input and filtered version, and thus drives up coherence of affected regions. So, improved filtering can help us potentially produce less biased coherence.

Next, we simulated 500 clean ground truth interferograms with Gaussian bubbles, roads and buildings, added Gaussian noise, and input noisy versions to each method mentioned in Fig. 4, including our method. The phce (mean of cosine of absolute phase error) and cmse (coherence mean squared error) between filtered and ground truth data are presented in Table I. We normalized ground truth and predicted coherence for each filter by subtracting their mean value before computing cmse scores.

TABLE I. RESULTS ON SIMULATED INTERFEROGRAMS

| Metric | *Boxcar* | *NLInSAR* | *Goldstein* | *NLSAR* | *Proposed* |
|---|---|---|---|---|---|
| **phce** | 0.9297 | 0.9153 | 0.8835 | 0.6647 | 0.8678 |
| **cmse** | 0.0283 | 0.0265 | N/A | 0.0813 | 0.0299 |

Only the NLSAR phce scores are better than ours, at the cost of much higher run time, fringe warping and phase flattening in noisy regions. For the cmse, our method's scores are very close to Boxcar and NLInSAR, although both of them create grainy coherence artefacts or spatial variance in coherence estimates, especially in incoherent regions. Thus, our CNN-based filtering and related coherence estimation (considered jointly) is a much faster solution which outputs better results than others, in terms of phase detail preservation and sharper, less grainy coherence.

### IV. CONCLUSION

We propose a CNN-based denoising filter and coherence estimation method for InSAR images. It preserves detail while denoising, and creates far less grainy / less specular coherence than existing methods. It demonstrates the capability of CNN-based learning for InSAR denoising and coherence estimation.